\documentclass[12pt]{iopart}

\usepackage{amssymb,amsfonts,bm}
\usepackage{graphicx}
\usepackage{hyperref}
\usepackage[dvipsnames]{xcolor}
\begin{document}

\title[Effect of increasing disorder on domains of the 2d CG]{Effect of increasing disorder on domains of the two-dimensional Coulomb glass}

\author{Preeti Bhandari$^{1}$ \& Vikas Malik$^{2}$}

\address{$^{1}$ Department of Physics, Jamia Millia Islamia, New Delhi 110025, India}
\address{$^{2}$ Department of Physics and Material Science, Jaypee Institute of Information Technology, Noida 201309, UP, India}
\ead{vikasm76@gmail.com}
\vspace{10pt}
\begin{indented}
\item[]9 August 2017
\end{indented}

\begin{abstract}
We have studied a two dimensional lattice model of Coulomb glass for a wide range of disorders at $T\sim 0$. The system was first annealed using Monte Carlo simulation. Further minimization of the total energy of the system was done using Baranovskii et al algorithm followed by cluster flipping to obtain the pseudo ground states. We have shown that the energy required to create a domain of linear size L in d dimensions is proportional to $L^{d-1}$. Using Imry-Ma arguments given for random field Ising model, one gets critical dimension $d_{c}\geq 2$ for Coulomb glass. The investigations of domains in the transition region shows a discontinuity in staggered magnetization which is an indication of a first-order type transition from charge-ordered phase to disordered phase. The structure and nature of Random field fluctuations of the second largest domain in Coulomb glass are inconsistent with the assumptions of Imry and Ma as was also reported for random field Ising model. The study of domains showed that in the transition region there were mostly two large domains and as disorder was increased, the two large domains remained but there were a large number of small domains. We have also studied the properties of the second largest domain as a function of disorder. We furthermore analysed  the effect of disorder on the density of states and showed a transition from hard gap at low disorders to a soft gap at higher disorders. At $W=2$, we have analysed the soft gap in detail and found that the density of states deviates slightly ($\delta\approx 1.293 \pm 0.027$) from the linear behaviour in two dimensions. Analysis of local minima show that the pseudo ground states have similar structure.
\end{abstract}

\vspace{2pc}
\submitto{\JPCM}
%
\maketitle
%
%

\section{Introduction}
\label{intro} The Coulomb glass (CG) is a system in which all electron states are localised and they interact via long-range Coulomb potential. At low temperature, these localised electrons are unable to screen the Coulomb interactions effectively. The long range nature of the Coulomb interactions leads to a soft gap \cite{Pollak,Sri,Shk} in the single-particle density of states (DOS). Efros and Shklovskii \cite{Shk} predicted a power law, $\rho(\varepsilon) \propto \vert \varepsilon-\mu \vert^{\delta}$ near the chemical potential $\mu$, with $\delta \geq d-1$ in d-dimensions. This effect changes the conductivity from $ln\sigma \sim T^{1/4}$ to $T^{1/2}$ as temperature ($T$) is decreased \cite{Mott1,Mott2}. The formation of gap and the crossover of $T^{1/4}$ to $T^{1/2}$ at low T have been confirmed experimentally and numerically \cite{Mob}. Another important effect of Coulomb interaction is correlation effects, i.e. existence of collective hops instead of single electron hops \cite{Kno}. In the recent years, focus has shifted from higher disorder to low disorder region \cite{Surer,Mobi}. Finite temperature simulations in three-dimensional ($3d$) CG  \cite{Goethe} have shown that a transition from fluid to the charge-ordered phase was consistent with the random field Ising model (RFIM) universality class.\\
\hspace*{3mm} One of the motivation of this paper is to understand the importance of Coulomb interactions in domain formation in the ground state and how the structure of the domain differs from the short range model i.e. RFIM. The Imry-Ma arguments \cite{Imry} on which the initial theoretical papers on RFIM \cite{Vill,Grin,Stauffer,Fisher} were based, suggested that the energy required for the formation of a large compact domain of linear size L in d-dimensions is $\mathcal{O}(L^{d-1})$. The amount of energy gained from the fluctuations of random field in the domain is $\mathcal{O}(L^{d/2})$, so the long range order will get destroyed for $d<2$. The ground state of 3d RFIM shows a transition from ferromagnetic to disordered state as disorder is increased \cite{Ogiel}. Binder \cite{Binder} argued that roughening of domain walls would stabilize the domain in two-dimensions and lead to destruction of ferromagnetic ordering. A rigorous proof was then given by Aizenman and Wehr \cite{Aizenman} stating that there is no long-range order in 2d RFIM. These arguments led to a critical dimension $d_{c}=2$. Numerical evidence \cite{Seppala,Seppala2} shows roughening of domain walls and the ground state breaking into domains above a length scale that depends exponentially on the random field strength squared, further strengthened the argument that $d_{c}=2$. Experiments on 2d dilute antiferromagnets, showed that no long-range ordering is present \cite{RJ2,IB1}, but a possibility of first order transition in 3d has been observed \cite{Cow}.\\ 
\hspace*{3mm} Contradicting all the above work, evidence of numerical signs of transition in 2d RFIM at $T=0$ below a critical disorder was shown by Frontera and Vives \cite{Frontera}. In 2013 Sinha and Mandal \cite{Suman} used Monte Carlo simulations to show that for weak fields 2d RFIM possesses long-range ordering. The validity of the Imry-Ma arguments in RFIM was tested by doing numerical calculations. The properties of domains were significantly different from the assumptions made by Imry and Ma \cite{Cambier,Esser}. In a completely compact domain, surface area ($S \sim L^{d-1}$) is related to the volume of the domain ($V \sim L^{d}$) by the relation $S \sim V^{\tau}$ where $\tau=d-1/d$. Cambier et al \cite{Cambier} found the value of $\tau=0.59$ for 2d and $\tau=0.84$ for 3d for disorder just above the critical disorder. For 2d since the value of $\tau>0.5$, they claimed that the domains were not compact (referred to as non compact in this paper). For 3d since $\tau$ is closer to $1$, they claimed that the domains are fractal. For 2d, at higher disorders Esser et al \cite{Esser} obtained value of $\tau=0.98$ and they also classified their domains as fractal. In our previous work \cite{Preeti}, we studied a 2d CG lattice model where we have shown a first order transition from charge-ordered phase to disordered phase at zero temperature using finite size scaling. According to our finite size analysis the critical disorder for $64\times64$ system was $0.265$. At critical disorder in contradiction with the Imry-Ma arguments, we found non compact domains where most of the random field energy was contained in the domain wall. Our results were inconsistent with the Binder's roughening arguments as well. A first order transition from liquid to stripe ordered phase has also been observed in $2d$ triangular lattice with Coulomb interactions \cite{Vladimir,Vla}. \\
\hspace*{3mm} In this paper we study the effect of increasing disorder on the domain of the $2d$ CG at $T\sim0$. We have first given a theoretical argument which shows that even for a system with long range Coulomb interactions the energy required to form a large completely compact domain is proportional to surface area of the domain in accordance with the Imry-Ma argument. It was verified numerically that this argument holds for large non compact domains also. We then studied the evolution of charge-ordered phase to a disordered phase as disorder was increased at zero temperature. We found that as disorder increased the state of the system changes from charge-ordered phase to disordered phase at a critical disorder ($W_{c}$). The effect of disorder on the domain structure was investigated and we found that at $W^{+}_{c}$ (i.e. disorder just above the critical disorder ($W_{c}$)), the second largest domain was non-compact and became fractal at large disorders. As disorder was increased, number of domains increases from 1 (at charge-ordered phase) to $\approx$2 at $W^{+}_{c}$ and a large number at high disorders. The structure and the number of domains also has an effect on DOS. We have then studied the effect of disorder on the DOS.\\
\hspace*{3mm} The rest of the paper is arranged as follows. In Sec.~\ref{models} we have discussed our model. In Sec.~\ref{NS} our minimization algorithm is discussed in detail. In Sec.~\ref{TA} we have given a theoretical argument that for a system with long range Coulomb interactions the energy required to form a regular domain scales as $L^{d-1}$ as suggested by Imry and Ma for short range interactions. Using our numerical data we have then shown that this argument holds for non compact domains also . In Sec.~\ref{OP} we are looking at the variation of order parameter around the critical disorder. In Sec.~\ref{PD} we have studied the properties of the domains as a function of disorder. In Sec.~\ref{DOS} we have investigated the DOS as a function of disorder. Also the gap exponent $\delta$ was calculated at higher disorders and compared with the earlier results. In Sec.~\ref{LMS} we have studied local minima statistics, where we found that the domains were pinned at a certain location independent of its initial state. The local minima (pseudo ground states) have similar structure. Conclusions are presented in Sec.~\ref{Conclu}.
\section{Model}
\label{models} We consider the classical 2d CG lattice model \cite{Shk}, in which the electron states are assumed to be localized around the sites of a square lattice with lattice spacing $a\equiv 1$. We work with a case of half filling which implies that the number of electrons are half the total number of sites (N). We use the pseudospin variables $S_{i}=n_{i}-1/2$ where $n_{i}\in {0,1}$ is the occupation number at site i. The Hamiltonian of the system can now be written in spin language as 
\begin{equation}
H = \frac{W}{2}\sum_{i} \phi_{i}S_{i} + \frac{1}{2} \sum_{i\neq j} J_{ij} S_{i}S_{j}
\end{equation}
where the unscreened Coulomb interactions are described as $J_{ij}=e^{2}/\kappa R_{ij}$, $\kappa$ is the dielectric constant and $R_{ij}$ is the distance between site i and j. We are using periodic boundary condition which eliminates the boundary effects and makes each site equivalent \cite{Davies,MR,glatz}. Hence the distance between sites $i$ and $j$ were considered as the length of shortest path. This limitation imposes a cut off in the Coulomb interactions at a distance equal to $L/2$ where $L$ is the linear dimension of the system. $\phi_{i}$'s denote the random on-site energies, chosen randomly from a box distribution with interval [-1,1] and $W$ is the disorder strength. The particle-hole symmetry with symmetric disorder distribution lead to $\mu=0$. All the energies were measured in the unit of $e^{2}/\kappa a$ and all distances in the unit of $a$.

\section{Numerical simulation}
\label{NS} We have done simulated annealing using Monte Carlo technique for $64\times 64$ square lattice. The initial system was completely random spin configuration $\lbrace S_{i}\rbrace$ with half sites assigned with $S_{i}=\frac{1}{2}$ and other half with $S_{i}=\frac{-1}{2}$. The $\lbrace \phi_{i}\rbrace$'s were chosen in the following manner: for each run, the configuration of signs of the random energies is kept fixed ($\lbrace \phi_{i}\rbrace$ was chosen from a box distribution $\lbrace -1,1 \rbrace$) and then multiplied by $W/2$ where $W$ is the disorder strength, which was increased from $0$ to $2.0$ in small steps (we have considered the following disorders: W=0.0, 0.1, 0.15, 0.18, 0.20, 0.21, 0.22, 0.235, 0.25, 0.265, 0.285, 0.30, 0.35, 0.40, 0.50, 0.65, 1.0, 2.0). Since the critical disorder for $L=64$ is $W=0.265$ \cite{Preeti}, we have chosen more points in the transition region. This approach has an advantage that one is able to see the evolution of domains as $W$ is increased. Annealing using Metropolis algorithm \cite{Metro} was done from $T=1$ to $T=0.01$ (where we have taken $T= 1$, $0.5$, $0.33$, $0.2$, $0.143$, $0.125$, $0.11$, $0.105$, $0.1$, $0.095$, $0.091$, $0.087$, $0.083$, $0.08$, $0.077$, $0.066$, $0.05$, $0.04$, $0.033$, $0.02$, $0.01$). At $W=0$, as temperature is decreased there is a transition from paramagnetic to charge ordered phase at $T_{c}=0.103$ \cite{mobb}. The critical temperature decreases as disorder is increased. Hence we have chosen a lot of temperatures between $T=0.143$ and $T=0.05$. At high disorders the Coulomb gap formation starts at $T=0.2$ and goes till $T=0.05$ \cite{Voj}. So the temperatures used for annealing are adequate for both low and high disorders used. Kawasaki dynamics was used as the number of electrons are conserved. A single electron hop attempt involves randomly choosing two sites of opposite spins for spin exchange. We used $3\times10^{5}$ number of Monte Carlo steps at high temperatures and   increased the number to a maximum of $5\times10^{5}$ at low temperatures. A single Monte Carlo step refers to $N/2$ spin exchange attempts.\\
\hspace*{3mm} The minimum energy state obtained from Monte Carlo simulation were then used for further minimization by an algorithm given by Baranovskii, Efros, Gelmont, and Shklovskii (we refer here by BEGS) \cite{begs}. The first criteria for minimization was that the Hartree energy of all electrons should be less then the Hartree energy of all holes. To do this subroutine called $``\mu-sub"$ given by BEGS was used in which Hartree energies ($\varepsilon_{i}$) \cite{Shk}
\begin{equation}
\varepsilon_{i} = \frac{W}{2} \phi_{i} + \sum_{j} \frac{S_{j}}{R_{ij}};
\end{equation}
of the minimum energy state was checked to see whether all occupied sites (i.e. $S_{i}=1/2$) had lower energies than the empty ones (i.e. $S_{i}=-1/2$). If this condition was not satisfied then, site with $S_{i}=1/2$ having the highest energy was exchanged with site having lowest energy and $S_{i}=-1/2$. The Hartree energies were then recalculated and the process was repeated until the ordering was correct. This was the end of $``\mu-sub"$. The second criteria for minimization was that the minimum energy state should be stable against all single electron hole transitions. So in the next step as mentioned by BEGS, we checked that the transition energy
\begin{equation}
\label{eqdel} \Delta^{j}_{i} = \varepsilon_{j} - \varepsilon_{i} - \frac{1}{R_{ij}}
\end{equation}
is positive for all pair of sites (consisting of opposite spins). If this condition was not satisfied by a pair, then their spins were exchanged and all the Hartree energies were recomputed and $\mu-sub$ was called again, followed by checking the condition in Eq.~\ref{eqdel}. This was repeated till both the conditions mentioned by BEGS were satisfied. The final local minimum energy state thus obtained is stable against all single electron hops and was termed as $``min\hspace*{1mm} state"$. The states obtained after Monte Carlo annealing were found to be very stable and hardly any minimization was achieved by using BEGS algorithm. The number of exchange done by BEGS algorithm were of the order of $2$ at higher disorders ($W>0.65$) and $\approx 0$ at lower disorders.\\
\begin{figure}
\centering
\vspace*{1cm}
\includegraphics[width=0.70\linewidth]{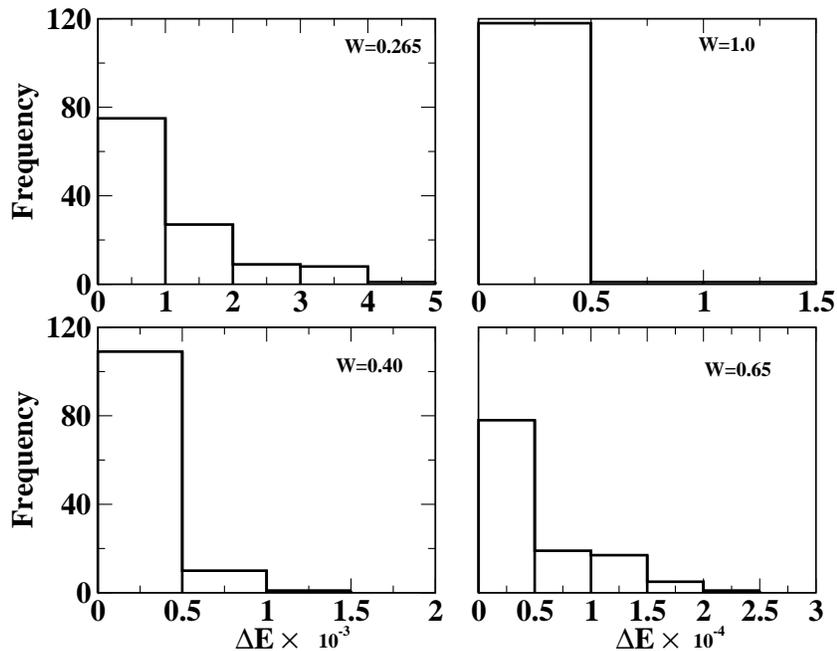}
\caption{(Color online) Relative energy difference between the pseudoground state after cluster flipping and the $min\hspace*{1mm} state$ before cluster flipping at different disorder strength.}
\label{Ediff1}
\end{figure}
\hspace*{3mm} The $min\hspace*{1mm} state$ was then used to perform cluster analysis by Hoshen-Kopelman algorithm \cite{Hoshen}. Using this algorithm we identified the domains in each configuration, where cluster of nearest -neighbour sites which have same $\sigma_{i}=(-1)^{i_{x}+i_{y}} S_{i}$ was defined as a domain where $i_{x}$ and $i_{y}$ are the $x$ and $y$ coordinates of site $i$ and $S_{i}$ is the spin at site $i$. The number of domains in each configuration was denoted by $n_{c}$. The domain-domain interaction was found to be negligible (excluding the interaction between the domain in question and the domain enclosing it ) for all domains. This allows one to flip domains independently of each other. Next a subroutine (delH) was called up. In this subroutine, a domain was flipped and the new energy of the system was calculated. The relative change in energy was calculated as
\begin{equation}
\label{eq2} \Delta E = \frac{E_{new} - E_{init}}{E_{new}}
\end{equation}
where $E_{new}$ is the energy of the system after domain flipping and $E_{init}$ is the energy of the initial state. Then the system was flipped back to its initial state. This procedure was repeated ($n_{c}-1$) times to calculate the energy required to flip each domain. The domain with minimum $\Delta E$ (where $\Delta E < 0$ only was considered) was selected among all the domains. This was the end of delH. The sites of the selected domain were then flipped. It is important to note that only those clusters were flipped which lowered the energy of the system. The flipped state was now considered as the initial state and the cluster identification was done again. Subroutine ``$delH$" was called up and the procedure was repeated till one got a single domain or the system reached the lowest energy state. The final state thus obtained was called the pseudoground state or metastable state. \\
\hspace*{3mm} In Fig.~\ref{Ediff1} we have shown the energy difference between the $min\hspace*{1mm} state$ and our pseudoground state at different disorder strengths. One can see that there is very small energy difference between $min\hspace*{1mm} state$ and our pseudoground state. But the structure of $min\hspace*{1mm} state$ and our pseudoground state is very different at $W < W_{c}$.The $min\hspace*{1mm} state$ at $W < W_{c}$ corresponds to a state with 2 large domains and few small domains and pseudoground state is a single domain state (charge-ordered phase). An important question is that, can one get the single domain state for small disorders using only annealing (without cluster flipping the final state got after annealing). One way is that for each disorder configuration $\lbrace \phi_{i}\rbrace$, anneal the system starting from different initial configurations $\lbrace S_{i}\rbrace$. Each initial configuration leads to a minimum energy state. The lowest energy state is then the pseudo ground state \cite{begs}. Another way is to anneal the system very slowly. Both these processes will be time consuming and therefore we have used cluster flipping. At higher disorders the domains are very stable and there is hardly any cluster flipping, so there is little difference between $min\hspace*{1mm} state$ and our pseudoground states. The minimum energy state (pseudoground state) is not the true ground state. In Sec.~\ref{LMS} we will show that the pseudoground states are similar to each other. In this paper, we have assumed our pseudoground state as the ground state and this is used throughout the paper unless mentioned otherwise.
\begin{figure}
\centering
\vspace*{1cm}
\includegraphics[width=0.40\linewidth]{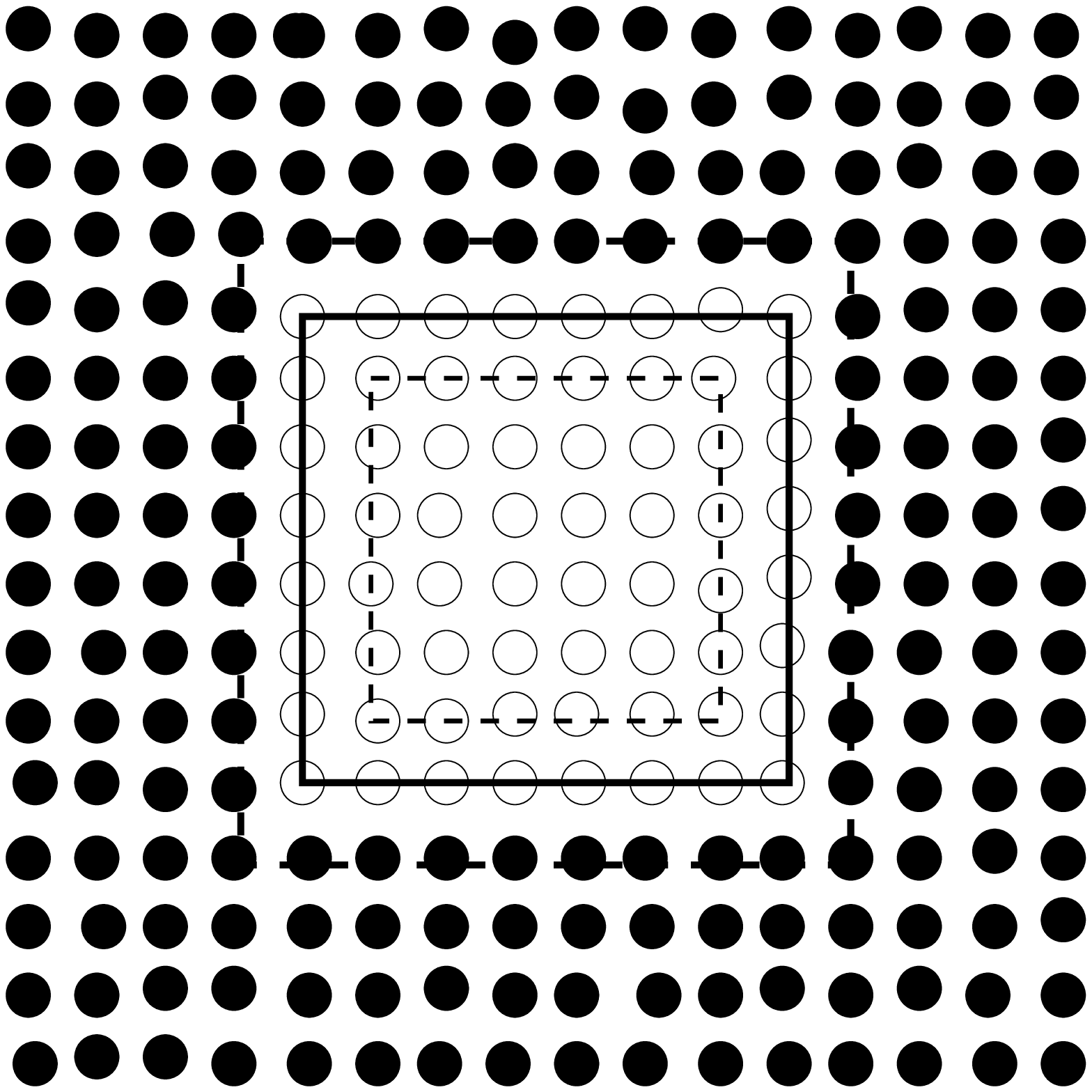}
\caption{(Color online) Filled circles denotes one antiferromagnetic phase and the empty circles denotes the other antiferromagnetic phase. The empty circles denote a square domain. Solid line denotes the domain wall, and the dashed line inside the domain wall and outside the domain denote the sites just inside and just outside the domain wall respectively. The figure not drawn to scale.}
\label{theorydomain}
\end{figure}
\section{Theoretical argument}
\label{TA} We are here proposing an argument to calculate the energy of a large regular domain which is square for $d=2$ (see FIG. \ref{theorydomain}) and cube for d=3, created in the ground state of a d-dimensional CG lattice model at half filling. The Hamiltonian of the system can also be written as
\begin{equation}
 H = \frac{1}{2} \sum^{N}_{i=1}(\varepsilon^{\prime}_{i} +2\phi_{i}) S_{i}
\end{equation} 
where $\varepsilon^{\prime}_{i} = \sum^{\prime}_{j} J_{ij} S_{j}$ is the interaction energy. The prime on summation sign here denotes that the term with $i=j$ is not considered. In the zero disorder limit, the ground state of a CG system has Antiferromagnetic ordering. So interaction energy at each site is equal to d-dimension Madelung energy ($\varepsilon^{'}_{d}$). Staggered magnetisation defined as $\sigma=1/N \sum^{N}_{i=1}\sigma_{i}$, is the order parameter. As the system has anti-ferromagnetic ordering, each row (in $2d$ and plane in $3d$) on the lattice is charge neutral. For any charge, the contribution to its interaction energy can be divided into two parts (a) from charges on the line (plane) on which the site is located (b) charges on few rows (planes) just above and below the charge under consideration and negligible contribution coming from rest of the lines(planes) \cite{Tasker}. There is no surface effect as we are using periodic boundary conditions. Now if we consider a large regular domain, which is square for $d=2$ (see FIG. \ref{theorydomain}) and cube for $d=3$, then the interaction energy of the sites inside the domain will be equal to $\varepsilon^{\prime}_{d}$ using the reasoning given above. For this argument, the planes we are considering are parallel to the surface of the domain. Extending the same argument, the interaction energy of the site on the surface of the domain (we call it domain wall) becomes approximately equal to $d-1$ Madelung energy ($\varepsilon^{\prime}_{d-1}$). This is because, for a site $i$ on the domain wall, $\varepsilon^{\prime}_{i}=\sum^{\prime}_{j} J^{F}_{ij} \sigma^{out}_{j} + \sum^{\prime}_{j} J^{F}_{ij} \sigma^{in}_{j} + \sum^{\prime}_{j} J^{F}_{ij} \sigma^{wall}_{j}$ where $J^{F}_{ij} = J_{ij} (-1)^{i_{x}+i_{y}+j_{x}+j_{y}}$ and $\sigma^{out}_{j},\sigma^{in}_{j},\sigma^{wall}_{j}$ describes $\sigma$ of sites outside the domain wall, inside the domain wall and on the domain wall respectively. Here the contribution from the first two terms in the summation is negligible, as the sites inside the domain wall cancels with the corresponding sites outside the domain wall. Rest of the sites are far away from the sites on the domain wall and have a negligible contribution to the site's energy. The third term of the summation is $\sim \varepsilon^{\prime}_{d-1}$ for a large regular domain. The above arguments will not hold for sites near the edges of the domain. So the energy required to create a domain wall is $\sim$ $((\varepsilon^{\prime}_{d}-\varepsilon^{\prime}_{d-1})/2)\times S$ where S is the surface area of the domain. Since energy gained from random field is still $\mathcal{O}(L^{d/2})$, Imry-Ma argument can be applied to a large regular domain. This might explain why 3d CG is in the same universality class of RFIM as claimed earlier \cite{Goethe}.\\
\begin{figure}
\centering
\vspace*{1cm}
\includegraphics[width=0.70\linewidth]{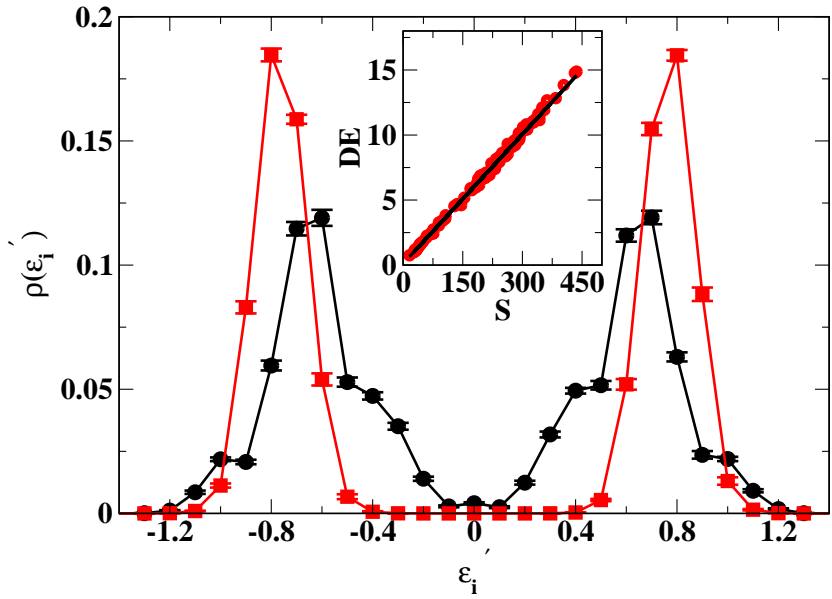}
\caption{(Color online) At $W^{+}_{c}$, distribution of interaction energy of the sites on the domain wall ($\varepsilon^{\prime \hspace*{1mm} wall}_{i}$) ($\fullcircle$) and on sites inside the domain wall where domain wall sites are excluded ($\varepsilon^{\prime \hspace*{1mm} inside}_{i}$) ($\fullsquare$). Inset shows the domain energy (DE) vs surface area (S) of the second largest domain.} 
\label{DEP}
\end{figure}
\hspace*{3mm} At $W^{+}_{c}$, we found that the second largest domain was non-compact (shown in Sec.~\ref{PD}). So one needs to numerically test the relationship $DE \propto S$ (where DE is the domain energy and S is the surface area of the second largest domain) for non-compact domain. The plot of DE vs S (FIG. \ref{DEP}) inset shows that the relation $DE = \eta \times S$ (where $\eta=0.033$) is valid here. The value of $\eta$ calculated numerically is slightly higher then the predicted theoretical value for 2d which is $\eta= (\varepsilon^{\prime}_{2d}- \varepsilon^{\prime}_{1d})/2 \approx 0.0285$.  To understand this variation in $\eta$ values, we have plotted the distribution of interaction energies of the sites on the domain wall ($\varepsilon^{\prime \hspace*{1mm} wall}_{i}$) and inside the domain wall ($\varepsilon^{\prime \hspace*{1mm} inside}_{i}$) in FIG. \ref{DEP}. The interaction energies of the sites on the domain wall and inside the domain wall are distributed symmetrically around $\varepsilon^{\prime}_{1d}$ and $\varepsilon^{\prime}_{2d}$ respectively. This is the reason why our numerical results are close to our argument given for DE calculation.
\begin{figure}
\centering
\vspace*{1cm}
\includegraphics[width=0.70\linewidth]{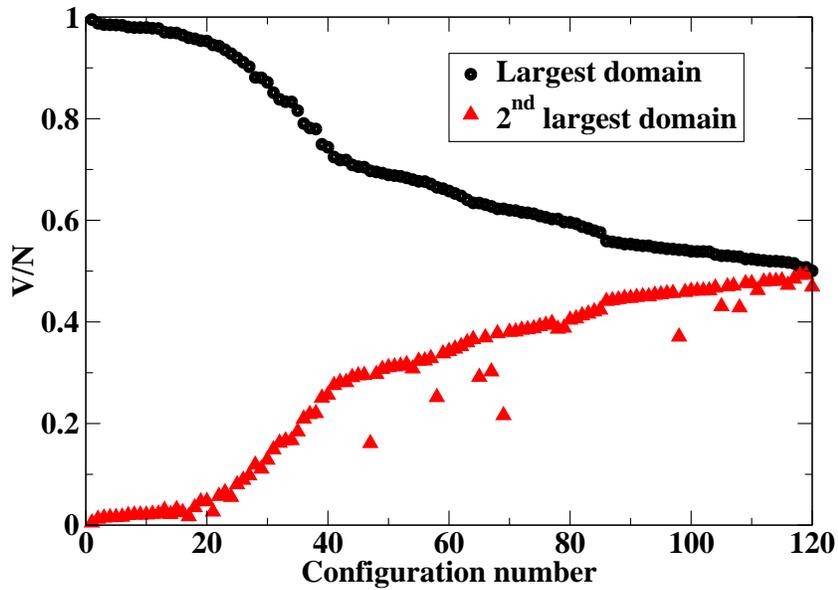}
\caption{(Color online) Largest and the second largest domain size ($V$) in the Disordered Phase ( at $W^{+}_{c}$) divided by the system size $N=64^{2}$. The largest domains obtained are sorted in descending order by size. The scatter in red dots corresponds to those configurations for which the charge-ordered phase breaks into more than two domains.} \label{domainBreak}
\end{figure}
\section{Order Parameter}
\begin{figure}
\centering
\vspace*{1.5cm}
\includegraphics[width=0.70\linewidth]{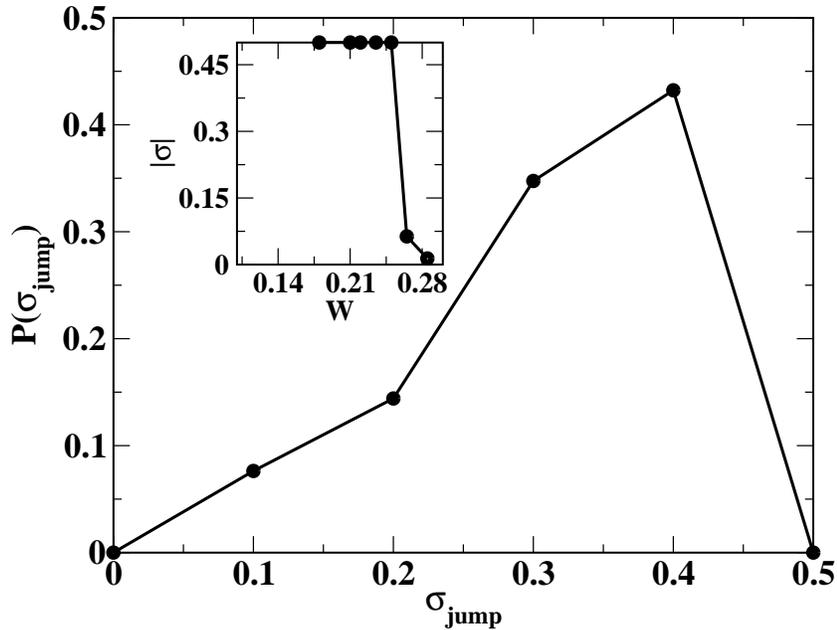}
\caption{(Color online) The probability distribution of largest jump in $\sigma$ at each configuration. Inset shows the jump in $\sigma$ for a single configuration.} \label{mjump}
\end{figure}
\label{OP} The investigations in this section were done at the transition region only ($W<0.30$) at zero temperature. At zero disorder and zero temperature, the system is in charge-ordered phase. As disorder is increased, for each disorder configuration (one set of $\lbrace \phi_{i}\rbrace$'s), the charge-ordered phase breaks into a multiple domain state. The disorder at which one gets a multiple domain state is denoted by $W^{+}_{c}$ and the maximum disorder at which charge-ordered phase exist is denoted by $W^{-}_{c}$. The critical disorder $W_{c}$ is different for different disorder configurations. This is finite size effect and dispersion in $W_{c}$ values would decrease as the system size increases. In FIG. \ref{domainBreak}, we have shown the size of the largest and the second largest domain for each configuration, where a single domain state (charge-ordered phase) at $W^{-}_{c}$ breaks into multiple domain state (disordered phase) at $W^{+}_{c}$. One can see that for most of the configurations charge-ordered phase breaks into two large domains as we move from $W^{-}_{c}$ (where $\sigma \hspace*{2mm} = \hspace*{2mm} \pm 0.5 $) to $W^{+}_{c}$ (where $\sigma \hspace*{2mm} = \hspace*{2mm} small$) which results into discontinuity in staggered magnetisation at each disorder configuration. The jump in $\sigma$ for one such configuration is shown in the inset of FIG. \ref{mjump}. This discontinuity in staggered magnetization is an indication of first order transition which was confirmed using finite size scaling approach in our previous paper \cite{Preeti}. In that paper, we have used $\lbrace \phi_{i}\rbrace$'s chosen randomly from a box distribution $[-W/2,W/2]$. There the critical disorder corresponds to that W at which maximum number of configurations make a transition from charge-ordered phase to disordered phase. In our present simulation, $W_{c}$ for each set of $\lbrace \phi_{i}\rbrace$ is different. From FIG. \ref{domainBreak}, one can see that in few configurations, size of the second largest domain is very small. In these configurations, the single domain state (where $\sigma=0.50$) first breaks into a small domain (where $\sigma \approx 0.4$) followed by a large jump in $\sigma$ (where $\sigma=small$). This was confirmed by the probability distribution for the largest jump in $\sigma$ at each disorder configuration as shown in  FIG.\ref{mjump}. Such small jumps in magnetisation and energy was also observed in 3d RFIM \cite{machta1,machta2}. These small jumps do not lead to a singular behaviour. Hence to look at thermodynamically favourable transitions one should analyse large jumps.
\section{Properties of the domain}
\label{PD} 
\begin{figure}
\centering
\vspace*{1cm}
\includegraphics[width=0.70\linewidth]{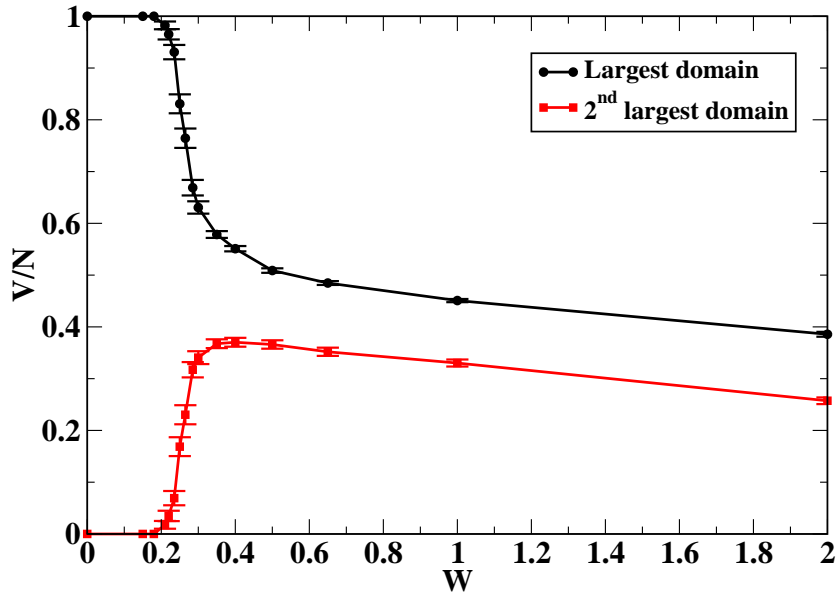}
\caption{(Color online) Behaviour of largest and the second largest average domain size ($V$) divided by the system size $N=64^{2}$ as a function of disorder. } \label{domainsize}
\end{figure}

\begin{figure*}
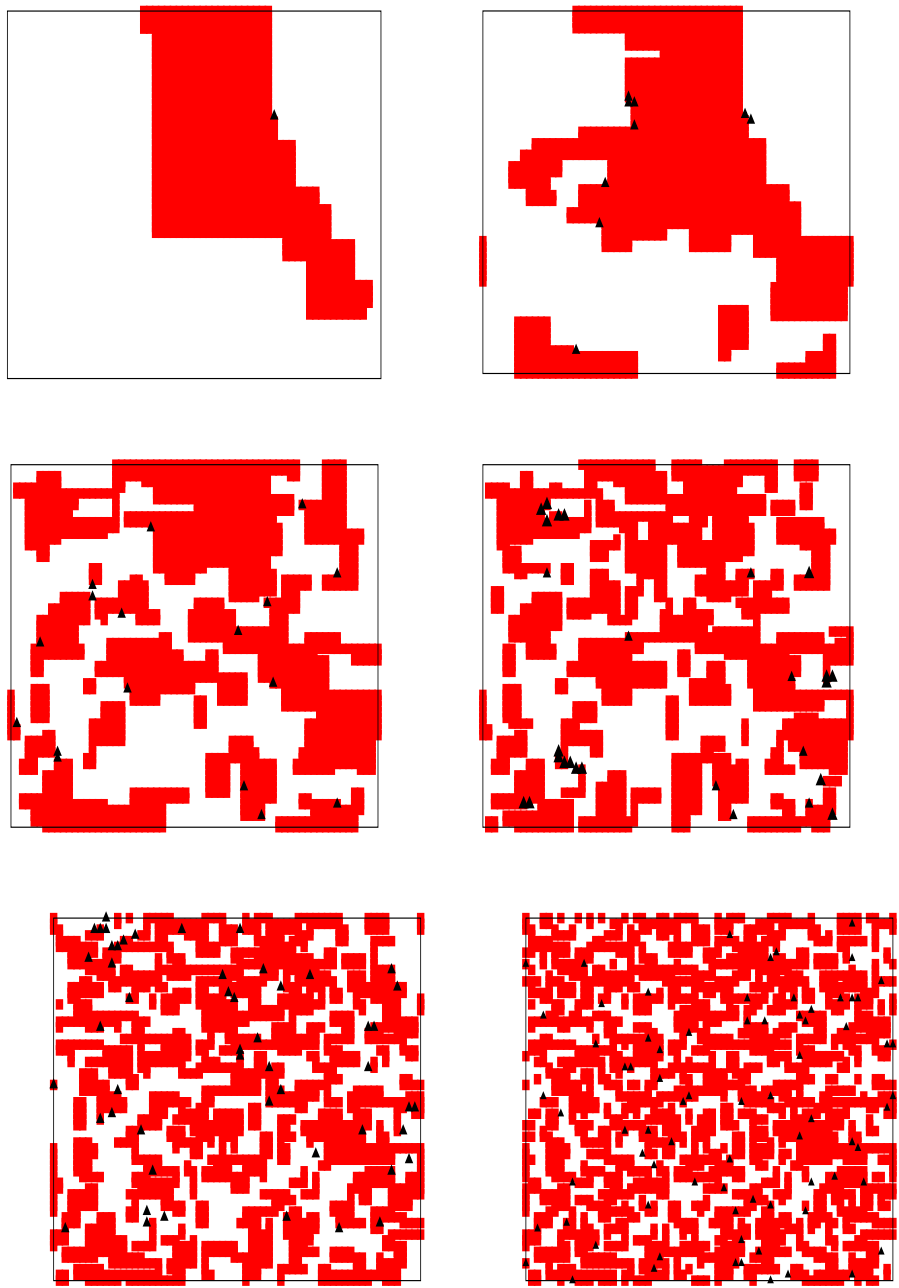

\centering
\vspace*{1cm}
\includegraphics[width=5cm,height=5cm]{Rfig7.eps}
\hspace*{1cm}
\includegraphics[width=5cm,height=5cm]{Rfig8.eps}
\hspace*{1cm}

\vspace*{1cm}
\centering
\includegraphics[width=5cm,height=5cm]{Rfig9.eps}
\hspace*{1cm}
\includegraphics[width=5cm,height=5cm]{Rfig10.eps}
\hspace*{1cm}

\vspace*{1cm}
\centering
\includegraphics[width=5cm,height=5cm]{Rfig11.eps}
\hspace*{1cm}
\includegraphics[width=5cm,height=5cm]{Rfig12.eps}

\caption{(Color online) Red and white regions denote different domains at $W=0.265,0.35,0.50,0.65,1.0,2.0$.$(\blacktriangle)$ are the sites which have Hartree energy in the range $-0.2 \leqslant \varepsilon_{i} \leqslant 0.2$} \label{domainStructure}
\end{figure*}

\begin{figure}
\centering
\vspace*{1cm}
\includegraphics[width=0.70\linewidth]{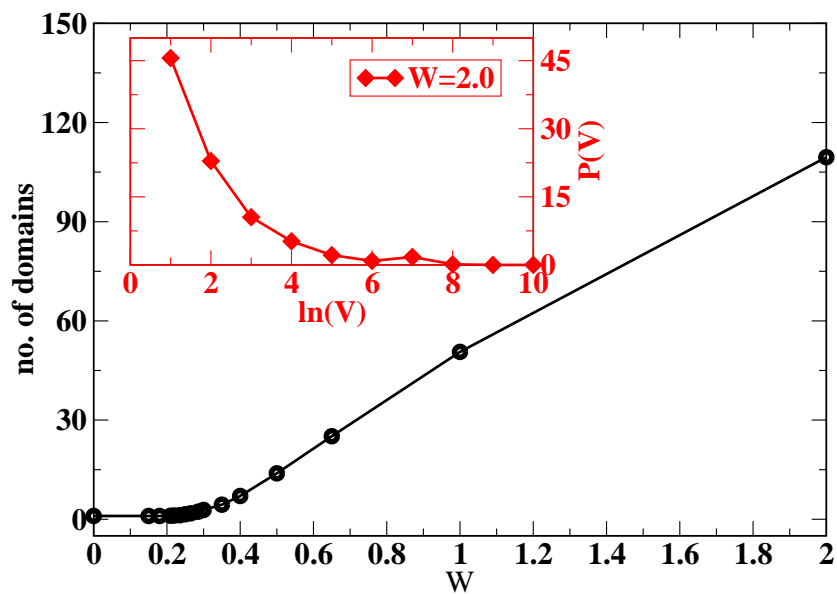}
\caption{(Color online) Number of domains as a function of disorder. In the inset we have shown the domain size ($V$) distribution of small domains at $W=2.0$ excluding the largest and the second largest domain.} \label{domainno}
\end{figure}

\begin{figure}
\centering
\vspace*{1cm}
\includegraphics[width=0.70\linewidth]{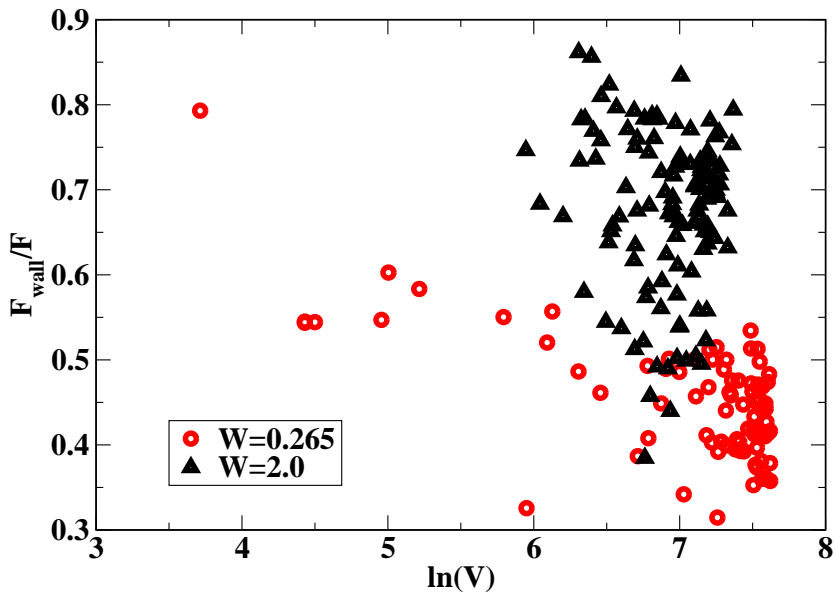}
\caption{(Color online) $F_{wall}/F$ vs $V$ at different disorders for the second largest domain.} \label{FwallF}
\end{figure}

\begin{figure}
\vspace*{1cm}
\centering
\includegraphics[width=0.70\linewidth]{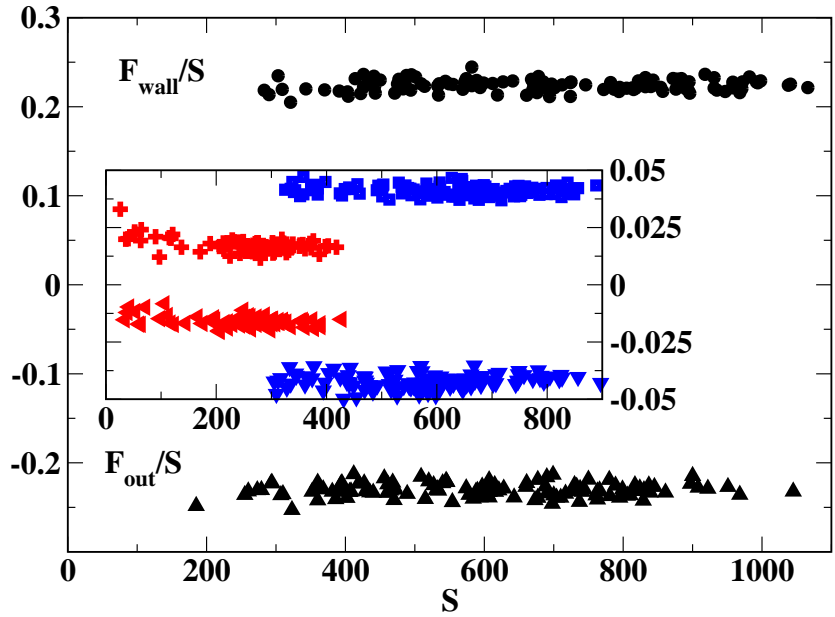}
\caption{(Color online) Random field fluctuation of the domain wall ($F_{wall}$) and just outside it ($F_{out}$) at $W_{c}=0.265$ (in red, $+$,$\blacktriangleleft$ ), $W=0.50$ (in blue, $\blacksquare$,$\blacktriangledown$) and at $W=2.0$ (in black, $\bullet$,$\blacktriangle$). The y coordinate is the ratio $F/S$, and the x coordinate is S.} \label{FwallP}
\end{figure}

In this section, we first looked at the variation in the size of the first and the second largest domain with disorder (as shown in FIG. \ref{domainsize}). We found that for $W<0.20$ one has charge-ordered state and then with increase in disorder the state breaks into two large domains. As disorder increases, the size of the first and second largest domain decreases. This is because of large number of small domains becoming stable at large disorder. The evolution of domains with disorder can be seen in FIG. \ref{domainStructure}. The distribution of number of domains with disorder is shown in FIG. \ref{domainno}. In the inset of FIG. \ref{domainno} we have shown the size distribution of domains (excluding the largest and second largest domains) that exist at $W=2.0$. One can see that the average size of the largest and second largest domains is less than the percolation threshold ($p_{c}=0.59$) for 2d \cite{Stauf}. \\
\begin{figure}
\centering
\vspace*{1cm}
\includegraphics[width=0.70\linewidth]{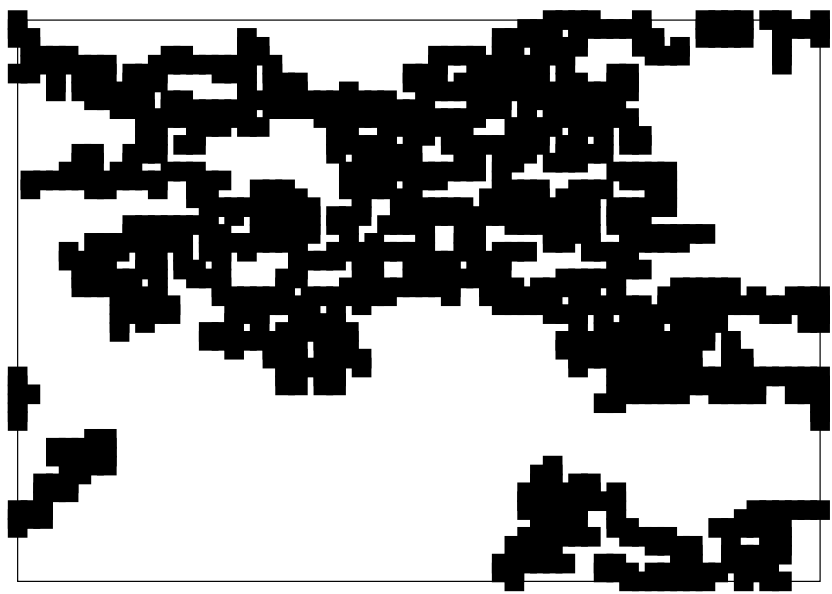}
\caption{(Color online) Second largest domain (black region) at $W=2.0$.} \label{Fractal}
\end{figure}
\hspace*{3mm} To test the validity of the Imry-Ma arguments on CG model, we focussed on the structure and the nature of random field fluctuations of the second largest domain at different W. The compactness of the second largest domain was checked by using a power law relation \cite{Cambier} $S\thickapprox V^{\tau}$, where the volume (V) of the domain is the total number of sites in the domain. When calculating the surface area and the volume of the second largest domain, we have not included contribution from holes (smaller domains inside the domain). The value of the surface exponent $\tau$ for a large compact domain is $1-1/d$. In our case a large square domain with no internal holes corresponds to a compact domain. The value of $\tau$ at different disorders is shown in Table \ref{T1}. The high value of $\tau$ at the transition region ($W=0.265$) indicates that the domains are non-compact. At $W=0.265$, we got mostly two large domains. The non-compactness of the second largest domain is due to roughening of the domain walls (effect due to any hole inside the domain has been neglected). The roughening corresponds to adjustment of the domain wall to minimize the total energy of the system. With increase in disorder $\tau$ increases as the non-compact second largest domain become fractal ($\tau \approx 1$). The second largest domain at $W=2.0$ is shown in FIG.\ref{Fractal} for the same configuration whose complete domain picture (including all the domains) is shown in FIG. \ref{domainStructure}. One can see that a non compact second largest domain at $W=0.265$ (FIG. \ref{domainStructure}) becomes fractal at $W=2.0$ (FIG.\ref{Fractal}). Next we tested the hypothesis that the total random-field fluctuations (F) in a domain is typically a rms deviation and is proportional to the square root of $V$. A general power law expression \cite{Cambier} can be written as $F\approx V^{\lambda}$ where $\lambda$ was considered as an undetermined exponent. The value of $\lambda$ at different disorders is summarized in Table \ref{T1}. One can see that even at low disorders $\lambda$ is significantly higher than the theoretical value ($\lambda=1/2$) given by Imry and Ma argument. \\
\hspace*{3mm} We have also calculated the ratio $F_{wall}/F$ at $W=0.265$ and at $W=2.0$ (high disorder region) at each configuration. In FIG. \ref{FwallF} one can see that the range of the ratio is $40\%$ to $60\%$ for most of the configurations at $W=0.265$ which increases to $60\%$-$80\%$ at $W=2.0$. This indicates that at $W=0.265$ the random field energy of the second largest domain is contained more in the domain boundary and this effect increases as the disorder increases. As shown in FIG. \ref{domainBreak} at the transition region ($W=0.265$), size of the second largest domain is small for some configurations. This is responsible for the dispersion in domain size at small disorder in FIG. \ref{FwallF}. But at high disorder this is not true. At high disorders, large number of small domains open up (as shown in FIG. \ref{domainno}) but two large domains are always present (see FIG. \ref{domainsize}). We then calculated the random-field fluctuation of the sites on the domain wall ($F_{wall}$) of second largest domain and of the sites which are just outside the domain wall ($F_{out}$) of second largest domain. FIG. \ref{FwallP} shows that the random-field fluctuations are proportional to the surface area of the domain (for all three disorder strengths) and not its square root as assumed in previous theories. One can see that $F_{wall}/S$ and $F_{out}/S$ are in the range $\pm (0.20 \hspace*{2mm} to \hspace*{2mm} 0.25)$ for $W=2.0$. This can be understood as follows: if all $S_{i}'s$ were anti-parallel to $\phi_{i}'s$ on the wall, then $\sum_{i} \phi_{i}S_{i}=-\frac{W_{avg}}{2}$, where $W_{avg}=\int^{W/2}_{0}W dW = 0.5$ for $W=2.0$. So one can see that domain wall of the second largest domain is following the random energy. To verify this point further, we have checked whether the spin orientation at the domain wall of second largest domain were aligned anti-parallel to the respective random field on that site or not. We found that the percentage of sites where the spin orientation is determined by the random field on that site increases as the disorder increases. At $W=0.265$, $60\%$-$70\%$ of the sites on the domain wall follow the random field which increases to $85\%$-$90\%$ for $W=2.0$.

\begin{table}
\centering
\caption{\label{T1} Structural exponents for the 2d Coulomb glass.}
\begin{indented}
\centering
\item[]\begin{tabular}{ @{} lll}
\br
$W$ & $\tau$ & $\lambda$  \\ 
\mr
0.265& 0.663 $\pm$ 0.018 & 0.655 $\pm$ 0.016 \\ 
0.50 & 0.773 $\pm$ 0.036 & 0.914 $\pm$ 0.019 \\ 
0.65 & 0.888 $\pm$ 0.038 & 0.942 $\pm$ 0.014 \\ 
1.0  & 0.923 $\pm$ 0.031 & 0.997 $\pm$ 0.010 \\ 
2.0  & 0.947 $\pm$ 0.047 & 1.000 $\pm$ 0.007 \\ 
\br
\end{tabular}
\end{indented}
\end{table}
\hspace*{3mm} So our results suggests that the second largest domain is non-compact in the transition region and becomes fractal as the disorder increases. Also the random field fluctuations are contained more at the domain boundary. These results are consistent with the numerical work \cite{Cambier,Esser} done on RFIM. As mentioned earlier, for each set of $\lbrace \phi_{i}\rbrace$, $W_{c}$ is different. So if one looks at the properties of domains at $W^{+}_{c}$, then $\tau=0.6756$ and $\lambda=0.6415$, which is close to what we are getting at $W=0.265$.

\section{Density of states}
\label{DOS}
\begin{figure}
\centering
\vspace*{1cm}
\includegraphics[width=0.70\linewidth]{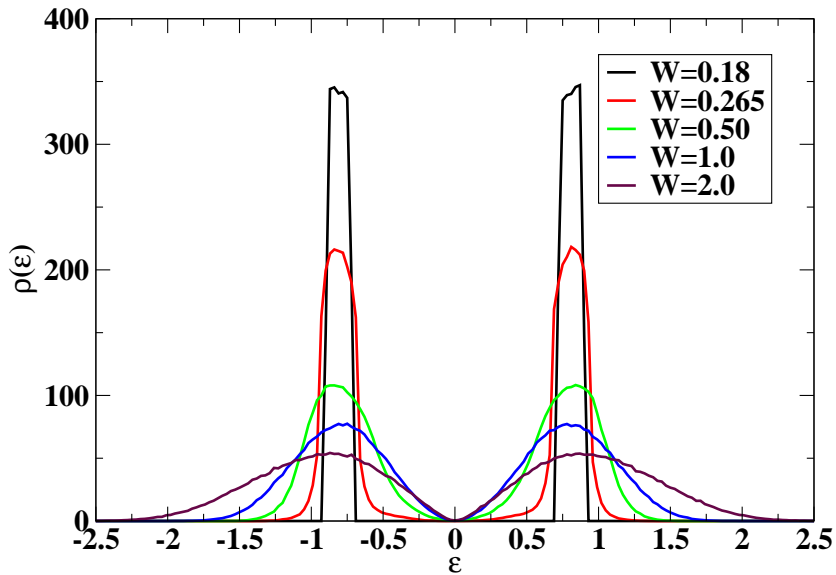}
\caption{(Color online)  Single-particle DOS for 2d in linear representation at different degree of disorder (W).}\label{dos1}
\end{figure}
\begin{figure}
\centering
\vspace*{1cm}
\includegraphics[width=0.70\linewidth]{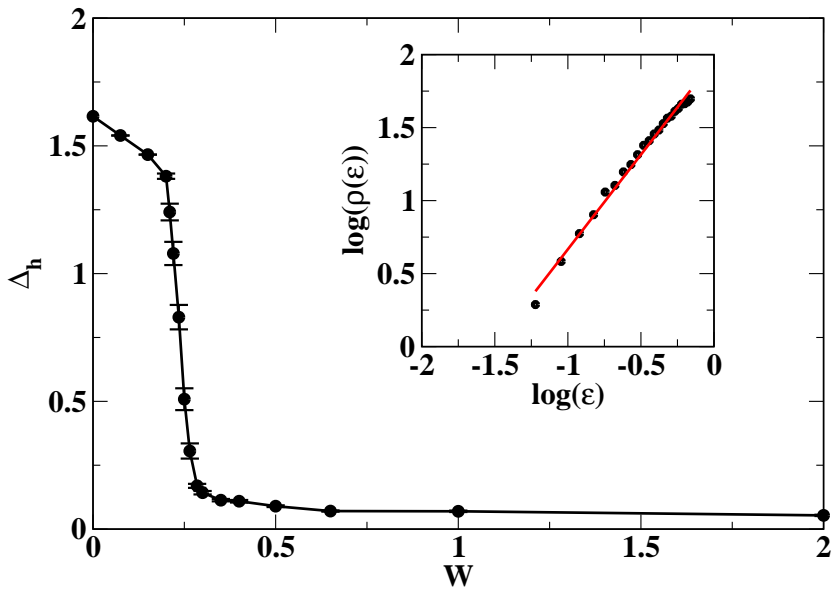}
\caption{(Color online) The hard Coulomb gap ($\Delta_{h}$) as function of disorder ($W$). (Inset) Averaged Coulomb gap at $W=2$ on a log-log scale. One can clearly see that the Coulomb gap deviates from the predicted linear behaviour $\delta=1.293 \pm 0.027$.} \label{dosgap}
\end{figure}

In 1975, Efros \cite{ef} used a self-consistent method to calculate DOS. He argued that in $2d$, DOS near the Fermi level calculated using single-particle excitation or polarons  (multi-particle excitations) are of same order i.e. $\rho(\varepsilon) \hspace*{2mm} \sim \hspace*{2mm} \mid \varepsilon \mid$. The first numerical simulation of the single particle DOS for 2d was performed by BEGS \cite{begs}. They considered a $16^{2}$ system with free boundary conditions at disorder strength $A=1$ ( which is $W=2$ in our notation). They calculated the DOS using pseudoground states which were stable against single electron hops. Their results show a linear behaviour in DOS ($\delta=1.0$) for the energy range $0.1 <\vert \varepsilon \vert \leq 0.7$ which was in agreement with the theoretical  predictions \cite{Shk}. For $\varepsilon<0.1$, their results were less accurate. Later Davies, Lee and Rice \cite{Davies} used a procedure very similar to BEGS but they carried out limited check for two electron hops also. They considered a $16^{2}$ lattice but with periodic boundary conditions. They have exploited the electron-hole symmetry and considered $\mu$ as the mean of the random energies $\phi_{i}$. We have used the same method to calculate $\mu$. They compared their work at $B=2$ with BEGS work at $A=1$ and showed that their results for $\rho(\varepsilon)$ were best fitted by a power law with an index of about $3/2$ near $\mu$. They attributed the change in $\delta$ mainly to the change in boundary conditions. A more sophisticated minimization techniques were adopted later \cite{MR,glatz}, which allows one to tackle very large system sizes. M\"{o}bius et al \cite{MR} found $\delta \simeq 1.2$ and Glatz et al \cite{glatz} found $\delta \simeq 1.23$ for 2d system. They also stressed that in 2d, there is a very small effect of stabilizing the ground state against two particle hops. We have also investigated the single-particle DOS at different disorders ($W$) as shown in FIG. \ref{dos1}. The gap exponent ($\delta$) at higher disorders was calculated and summarized in Table \ref{T2}. Comparing our results at $W=2$ (scaling shown in the inset of FIG. \ref{dosgap}) with the previous work \cite{begs,Davies,MR,glatz} we also found DOS deviating from the linear behaviour ($\delta \simeq 1.293 \pm 0.027$).\\
\hspace*{3mm} Glatz et al have studied the effect of disorder on DOS for $256\times256$ square lattice. They used $U\alpha_{i}$ with $\alpha_{i}\hspace*{1mm}\epsilon\hspace*{1mm}[-1,1]$ defined as the quenched uniformly distributed random site energies, where $U$ is their disorder strength. They defined a hard Coulomb gap as
\begin{equation}
\label{eqdelta} \Delta_{h} \equiv \varepsilon^{(0)}_{min} - \varepsilon^{(1)}_{max} ;
\end{equation}
where $\varepsilon^{(0)}_{min}$ was the minimum Hartree energy of the unoccupied site and $\varepsilon^{(1)}_{max}$ was the maximum Hartree energy of the occupied site. Below $U_{c}\approx 1/4$ (which is $W=0.50$ in our notation) they found a charge-ordered phase, where $\Delta_{h}$ decreases linearly with U. At $U_{c}$ they found a sharp discontinuity in $\Delta_{h}$  as they claim that the global crystalline structure breaks down and the hard gap disappears ( $\Delta_{h}\approx0$). This is attributed as a sharp transition from charge-ordered phase to probably a glassy phase. We have also calculated the hard Coulomb gap ($\Delta_{h}$) as defined by Glatz et al (in Eq. \ref{eqdelta} ) and its behaviour as function of $W$ is shown in FIG. \ref{dosgap}. One can see, that $\Delta_{h}$ shows a sharp change at $W=0.265$. This corresponds to a transition from charge-ordered phase to disordered phase at $W \approx 0.265$. From our previous work \cite{Preeti} we know that as $L$ increases, $W_{c}$ decreases towards $W=0.2253$ and the transition becomes sharper. Therefore one expects for larger system sizes, $\Delta_{h}\approx0$ will occur at lower disorder (tending towards $W=0.2253$) . At $W \approx 0.35$, $\Delta_{h} \approx 0$. The reason for $\Delta_{h} \approx 0$ is that some of the sites on the edges of the domains have very low energy. In FIG. \ref{domainStructure}, we have shown sites having small Hartree energies ($-0.2 \leqslant \varepsilon_{i} \leqslant 0.2$). The number of such sites increases as the domains become more non-compact leading to a soft gap at $W=0.50$.

\begin{table}
\centering
\caption{\label{T2} Gap exponent($\delta$) at different $W$ for the 2d Coulomb glass.}
\begin{indented}
\centering
\item[]\begin{tabular}{ @{} ll}
\br
$W$ & $\delta$  \\ 
\mr
0.50&1.974  $\pm$ 0.075 \\ 
0.65&1.809 $\pm$ 0.042 \\ 
1.0&1.604 $\pm$ 0.020 \\ 
2.0&1.293 $\pm$ 0.027 \\ 
\br
\end{tabular}
\end{indented}
\end{table}
\section{Local minima statistics}
\label{LMS} It is known \cite{Kogan} that different initial random configuration $\lbrace S_{i}\rbrace$ yield different energy minima at higher disorders. The lowest energy state is assumed as a ground state and the rest of the higher energy states are considered as the pseudo ground states or metastable states. We have here studied the distribution of these local minima states at different disorder strengths. Starting with 240 different initial random configuration $\lbrace S_{i}\rbrace$ for a single disorder configuration $\lbrace \phi_{i}\rbrace$ at $W=0.265$ to $W=2.0$, we carried out our minimization algorithm as mentioned in Sec.~\ref{NS}. For $240$ different initial configurations we obtained $240$ different pseudo ground states. The fact that each initial distribution leads to a different pseudo ground implies that the pseudo ground states obtained are a small fraction of the total number of pseudo ground states present in the system. We found that all the pseudo ground states obtained at each $W$ have similar domains structure and are pinned at a certain location. At each disorder, we have then calculated the number of active sites between all the obtained pseudo ground states .i.e. the number of spins exhibiting $S^{\alpha}_{i} \neq S^{\beta}_{i}$, where $\alpha$ and $\beta$ represent two different pseudo ground states. The number of active sites divided by the total number of sites (N) were defined as \cite{glatz}
\begin{figure}
\centering
\vspace*{1cm}
\includegraphics[width=0.70\linewidth]{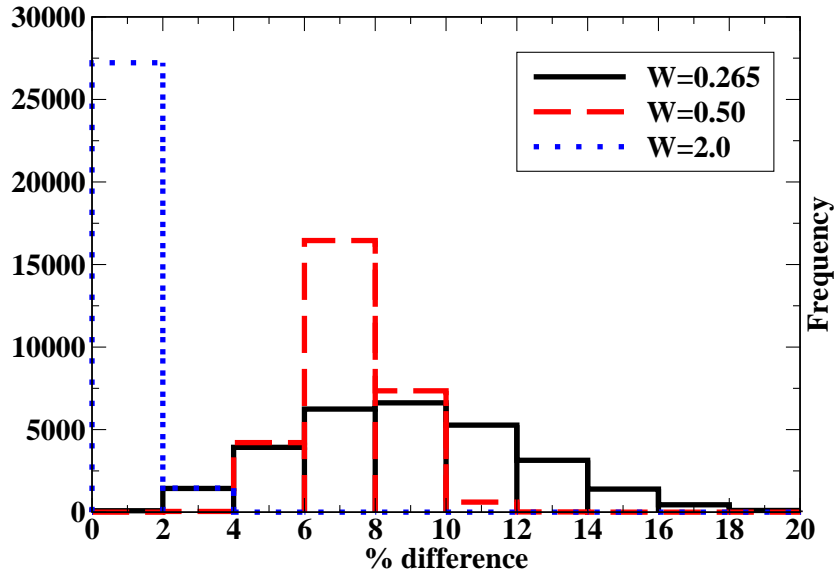}
\caption{(Color online) Distribution of the percentage difference in occupancy between pseudo ground states at each W.} \label{occdiff}
\end{figure}
\begin{equation}
\Delta_{\alpha\beta} = \frac{1}{N}\sum^{N}_{i} (S^{\alpha}_{i} - S^{\beta}_{i})
\end{equation}
The distribution of $\Delta_{\alpha\beta}$ (for all pairs of $\alpha,\beta$) at each disorder is shown in FIG. \ref{occdiff}. One can see that around the transition region, $\Delta_{\alpha\beta}$ are high which decreases as one increases the disorder strength. This is because around the transition region many domain wall rearrangements are possible (with very small energy difference) leading to high number of excitations. Such a situation was also observed in 3d RFIM \cite{Hart}. At higher disorders spin orientation are dominated by the $\lbrace \phi_{i}\rbrace$, which is fixed for all configurations at each $W$ here. Hence the number of active sites reduces drastically. From FIG. \ref{occdiff} one can see that even at low disorders the percentage difference in occupancy between different pseudo ground states in less than $20\%$. The energy difference between these states is $\mathcal{O}(1)$. Hence one can assume that the properties of these pseudo ground states are not very different from each other.
\section{Conclusions}
\label{Conclu} We have shown that the Imry-Ma argument for short range RFIM can be extended to CG system at half filling. We have numerically investigated two dimensional CG lattice model using Monte Carlo annealing to very low temperature from zero to large disorder. The energy of the annealed state was further minimized using BEGS algorithm and cluster flipping to obtain the pseudo ground states. We then investigated the domains around the critical disorder and found a transition from charge-ordered phase to disordered phase, leading to discontinuity in the staggered magnetisation as reported in our earlier work. The evolution of disordered phase with increasing disorder was studied next. We found that the second largest domain formed in the disordered phase is non-compact in the transition region and fractal at higher disorders. The random field fluctuations are contained more at the domain wall of second largest domain in contradiction with Imry-Ma assumptions. We have also investigated the effect of disorder on DOS. Our analysis of DOS at $W=2$ shows that the gap exponent $\delta\approx 1.293 \pm 0.027$, which is greater than the theoretical predictions \cite{Shk} but close to the previous numerical results \cite{Davies,MR,glatz}. An important question is what is the effect of multi-particle transitions on the nature of the ground state at high disorders. Our pseudoground state is stable against single particle transitions and cluster flipping. At high disorders the cluster flipping was found to be unimportant. It is possible that a very different minimum energy state could be obtained if multi-particle transitions were allowed during the annealing. This could lead to DOS with gap exponent closer to one as theoretically argued by Efros. It will be interesting to study the domain structure of this minimum energy state. The absence of hard gap coincides with our critical disorder which is much smaller than the value reported by Glatz et al. Filling of the gap at intermediate disorders is due to increase in non-compactness of the domains with increase in disorder. 
\section{Acknowledgements}
We thank late Professor Deepak Kumar for useful discussions on the subject. We wish to thank NMEICT cloud service provided by BAADAL team, cloud computing platform, IIT Delhi for the computational facility. P. B. acknowledges the University Grants Commission (UGC), India for financial support through Basic Scientific Research fellowship.

\end{document}